\documentstyle[prcrc,11pt,fleqn]{article}

\title{Baryon/anti-baryon inhomogeneity and \\ big bang nucleosynthesis}

\author{Kevork Abazajian and George M. Fuller 
\address{Department of Physics,
        University of California, San Diego, \\
        La Jolla, California 92109-0319}}

\begin{document}
\maketitle

\begin{abstract}
We investigate the effects of baryon/anti-baryon inhomogeneity on 
primordial nucleosynthesis. Recent work claims that electroweak 
baryogenesis could give rise to distinct regions of net baryon and 
anti-baryon number, which could survive until the nucleosynthesis epoch. 
We discuss neutron diffusion effects on nucleosynthesis yields in these 
models and speculate on the prospects for obtaining constraints.
\end{abstract}

\section{Introduction} Big bang nucleosynthesis (BBN) has been one of the
most successful enterprises of modern cosmology.  Comparison of
observationally inferred primordial light element abundances with
predictions from theoretical BBN calculations provide strong constraints
on conditions of the Early Universe.  The Early Universe somehow must
either produce the simple and appealing homogeneous conditions of the
standard BBN model, or it must produce an exceptional set of inhomogeneous
conditions that generate the observed light element abundances.  Prompted
by work on the QCD phase transition, the effect of high-amplitude,
sub-horizon scale, baryon-number fluctuations on BBN were studied
\cite{AHS,AFM,FMA}.  A detailed numerical calculation of the evolution of
such inhomogeneities through nucleosynthesis was done by Jedamzik, Fuller,
and Mathews \cite{JFM94}.  Neutrino-, baryon-, and photon-induced
dissipative processes were coupled to nuclear reaction rates in multiple
regions in an extension of the standard Wagoner \cite{WFH} nucleosynthesis
code.  In Ref. \cite{JFM94}, it was found that any significant
inhomogeneities overproduced helium-4 and/or deuterium.  The effect of a
first-order electroweak transition that produced similar baryon-number
inhomogeneities has also been studied \cite{FJMO94}.

Recently, Giovannini and Shaposhnikov \cite{GS} have suggested that
electroweak baryogenesis could take place through baryon-number-violating
interactions with a primordial (hyper-)magnetic field.  The primordial
field could fluctuate across regions much larger than the horizon at the
electroweak epoch, and also could cause both positive and negative
baryon-number fluctuations.  This could leave baryon/anti-baryon
homogeneities that conceivably could survive until the BBN epoch.  In Ref.
\cite{GS}, constraints were placed on the fields that guaranteed that they
would have no effect on standard BBN.  That is, the baryon/anti-baryon
regions were constructed to be smaller than the baryon diffusion length at
the epoch of BBN, so that such regions were damped out early. These
fluctuations were constructed so that the weak interactions reset the
resulting neutron-to-proton ratio, $n/p$, to its standard BBN value.

The $n/p$ ratio is extremely important in determining the resulting
element abundances.  Since almost all neutrons are eventually
incorporated into alpha particles, any change in the $n/p$ ratio will
affect the $^4$He abundance.  Roughly, the mass fraction, $Y_p$, of
$^4$He will be 
\begin{equation}
Y_p \approx \frac{4 n_4}{n_n + n_p} = \frac{2 (n/p)}{(n/p)} \approx 1/4.
\end{equation}
Baryon/anti-baryon regions that are larger than the baryon 
diffusion length could significantly affect the $n/p$ ratio.  Rehm and 
Jedamzik \cite{RJ} have recently studied these issues.

\section{Nucleosynthesis with baryon/anti-baryon 
inhomogeneities}

Baryon and anti-baryon regions that survive until the neutron--proton weak
interaction freeze-out epoch subsequently will face neutron/antineutron
diffusion and, therefore, baryon/anti-baryon annihilation. Ultimately,
that which we call ``matter"  must win out in annihilations and thus must
have a greater net number than antimatter.  Weak freeze-out occurs at a
temperature $T_{WFO} \sim 1~\rm{MeV}$ when neutrons and protons (and their
antiparticles) cease to be rapidly interconverted one to another by lepton
capture processes.  Separate regions of baryons or anti-baryons will be
preserved for epochs $T > T_{WFO}$ if the length scales associated with
these fluctuations are large compared to the baryon (anti-baryon) 
diffusion length. However for $T < T_{WFO}$, neutrons and antineutrons
will diffuse efficiently between these regions, while protons and
antiprotons remain relatively fixed in their respective regions because of
coulomb interations with the $e^{\pm}$ background.  Neutrons and
antineutrons are free from such interactions and their diffusion length
is several orders of magnitude larger.  Neutrons ``free stream" into
anti-baryon regions, and antineutrons ``free stream" out into the baryon
regions.  Assuming the baryon regions are significantly larger (in total
number of baryons), annihilations between neutrons and anti-baryons in the
former anti-baryon regions have two important effects:  (1)  depletion of
neutrons in the baryon regions, and (2) the formation of extremely
neutron-rich ``bubbles" in place of the anti-baryon regions. 

Recent work on the observationally inferred primordial abundances of the
light elements D, $^4$He, and $^7$Li have left a slight, yet significant
disparity between the predicted standard BBN abundances and measurements
\cite{BT,OSS,BM}.  Observationally inferred abundances and corresponding
BBN-inferred baryon-to-photon ratios ($\eta$'s) are: 
$Y_p  =  0.234~\pm~0.002,
\eta_{\rm{He}}  =  (1.8~\pm~0.3) \times 10^{-10}; 
{\rm{D}}/{\rm{H}}  =  (3.4~\pm~0.3~\pm~0.3) \times 10^{-5}, 
\eta_{\rm{D}}  = (5.1~\pm~0.3) \times 10^{-10};
{^{7}\rm{Li}}/{\rm{H}} =  (3.2~\pm~0.12~\pm~0.05) \times 10^{-10},
\eta_{\rm{Li}} = (1.7^{+0.5}_{-0.3}) \times 10^{-10} 
~\rm{or} ~(4.0^{+0.8}_{-0.9}) \times 10^{-10}$.
Specifically, the production of $^4$He is inconsistent (too high) with
the $\eta$ inferred from the primordial deuterium abundance.

Neutron-depleted baryon regions and the neutron bubbles have interesting
consequences for BBN.  The depletion of neutrons lowers the $n/p$ ratio 
and, therefore, lowers the predicted $^4$He abundance.  We can show that
the number density, $n_{\bar{b}}$, of the anti-baryon regions that occupy
a
fraction of the horizon volume, $f_v$, is related to the number density of
the baryons, $n_b$, in the baryon regions by 
\begin{equation} 
n_{\bar{b}} = \left[ \frac{k_{\rm{NSE}} (1-f_v) - k_{\rm{He}}}{f_v
(1-f_v+k_{\rm{He}})} \right]  n_b . 
\end{equation}
Here, $k_{\rm{NSE}}$ is the standard BBN $n/p$ ratio, and $k_{\rm{He}}$ is
the $n/p$ ratio that corresponds to the observed $^4$He abundance. The
number densities $n_b$ and $n_{\bar{b}}$ refer to the epoch just prior to 
$T_{WFO}$.  If the initial
number density of the anti-baryon regions is equal to the number density
of
the baryon regions, then the fractional volume that can be occupied by
anti-baryons for concordance between $^4$He and observationally
inferred deuterium abundance is
$f_v = 10^{-4}$. This small fraction reflects the small order of disparity
between the light element abundances.  Anti-baryon regions with
slightly larger $f_v$ will underproduce $^4$He; anti-baryon regions with
significantly larger $f_v$ will not have all of their anti-baryons
annihilated prior to nucleosynthesis.  
  
Nucleosynthesis in the neutron-rich regions is also important when
considering baryon and anti-baryon inhomogeneities.  In order to simulate
nucleosynthesis in these regions, we have forced the weak interactions
rates to freeze-out the $n/p$ ratio at $\sim 6$ and $\sim 10$. The
``isospin symmetric" nucleosynthesis of $n/p \sim 6$ would be expected
to evolve identically to the standard case of $n/p \sim 1/6$, with
nucleosynthesis being limited by the lack of protons rather than
neutrons.  However, the decay of neutrons provides a source for protons
in these neutron-rich regions and can lead to unique nucleosynthesis
results.  In our numerical calculations, the production of beryllium-9
and boron-11 become significantly amplified.  The recent limits on the
abundance of these elements in low metallicity stars provide a strong
upper bound on their primordial production \cite{Dunc,Mol}.

We have also included reactions leading to heavier element 
production in neutron-rich regions:
\begin{equation}
^{14}\rm{C}(\alpha,\gamma)^{18}\rm{O}(n,\gamma)^{19}\rm{O}(\beta)
^{19}\rm{F}(n,\gamma)^{20}
\rm{F}(\beta)^{20}\rm{Ne}(n,\gamma)^{21}\rm{Ne}(n,\gamma)^{22}\rm{Ne} .
\end{equation}
Through further $(n,\gamma)$ reactions and $\beta$-decays, the products
of these reactions may trigger the r-process in the neutron-rich
bubbles.  Primordial production of heavy elements is also constrained
by older Population II stars.  However, a limited primordial source of
heavy elements can not be ruled out. 

Whether or not values of the parameters $n_b$, $n_{\bar{b}}$, and
$f_v$ exist for which all of the nucleosynthesis constraints can be
met remains an open question.  It appears to us, however, that it
will be extremely difficult to find such values, especially in light
of potential B, Be, and r-process production in these schemes.  

\section{Acknowledgment}
This work was supported in part by grants from NSF and from NASA.


\begin{thebibliography}{9}
\bibitem{AHS} J. H. Applegate, C. J. Hogan, and R. J. Scherrer, 
ApJ 329 (1997) 572.

\bibitem{AFM} C. Alcock, G. M. Fuller, and G. J. Matthews, 
ApJ 320 (1987) 439.

\bibitem{FMA} G. M. Fuller, G. J. Mathews, and C. R. Alcock, 
Phys. Rev. D 37 (1988) 1380.

\bibitem{JFM94} K. Jedamzik, G. M. Fuller, and G. J. Mathews,  
ApJ 423 (1994) 50.

\bibitem{WFH} R. V. Wagoner, W. A. Fowler, and F. Hoyle, 
ApJ 148 (1967) 3.

\bibitem{FJMO94} G. M. Fuller, K. Jedamzik, G. J. Mathews, and A.
Olinto, Phys. Lett. B 333 (1994) 135.

\bibitem{GS} M. Giovannini and M. E. Shaposhnikov, 
Phys. Rev. Lett. 80 (1998) 22.

\bibitem{RJ} J. B. Rehm and K. Jedamzik, astro-ph/9802255 (1998).

\bibitem{BT} S. Burles and D. Tytler, astro-ph/9712265 (1997).

\bibitem{OSS} K. A. Olive, G. Steigman, and E. D. Skillman,  
ApJ 483 (1997) 788.

\bibitem{BM} P. Bonifacio and P. Molaro,  
Mon. Not. R. Astron. Soc. (1997) 847.

\bibitem{Dunc} D. K. Duncan, L. M. Rebull, F. Primas, et al.,
Astron. Astrophys. 332 (1998) 1017.

\bibitem{Mol} P. Molaro, P. Bonifacio, F. Castelli, and L. Pasquini, 
Astron. Astrophys. 319 (1997) 593.

\end{thebibliography}
\end{document}